\documentclass[12pt,preprint]{aastex}

\newcommand{\etal}{et al.\ }
\newcommand{\kms}{\, {\rm km\, s}^{-1}}

\newcommand{\ergs}{\, {\rm erg} \, {\rm s}^{-1}}
\newcommand{\cm}{\, {\rm cm}}

\newcommand{\kpc}{\, {\rm Kpc}}

\newcommand{\msun}{M_{\odot}}
\newcommand{\lya}{Ly$\alpha$ }
\newcommand{\hi}{\mbox{H\,{\scriptsize I}\ }}
\newcommand{\heii}{\mbox{He\,{\scriptsize II}\ }}

\slugcomment{Submitted to ApJL}

\shorttitle{Self-Absorption in QSO's}

\shortauthors{Alam \& Miralda-Escud\'e}

\begin{document}

\title{Self-Absorption of Ionizing Radiation and Extended Narrow-Line
Emission in High-Redshift QSOs}

\author{S. M. Khairul Alam\altaffilmark{1} \&
Jordi Miralda-Escud\'e \altaffilmark{1,2}
}
\altaffiltext{1}{Department of Astronomy, The Ohio State University, 
Columbus, OH 43210; alam,jordi@astronomy.ohio-state.edu}
\altaffiltext{2}{Alfred P. Sloan Fellow}

\begin{abstract}
  We calculate the neutral hydrogen column density of self-absorption in
QSOs predicted in a model where the QSOs are located in the same halos
that contain the gas in damped \lya absorption systems. The model is
parameterized by the probability $P_0$ that any halo has an active QSO.
We assume that the QSOs ionize the gas, but do not expel or heat it. The
derived \hi column densities produce negligible Lyman limit absorption,
even in the lowest luminosity QSOs, with an optical depth of only $\sim
10\%$ for luminosity $L=0.01 L_*$, when $P_0=10^{-2}$. We also compute
the \heii Lyman limit self-absorption, which is slightly higher but
still negligible. The self-absorption can be higher if the gas is highly
clumped; only in this case the overall emissivity from QSOs could be
significantly reduced due to absorption by the known damped \lya
systems, to affect the predicted intensity of the ionizing background or
the epoch of \heii reionization. The presence of the gas associated with
damped absorption systems around QSOs could also be detected from the
narrow \lya emission line, which should have and angular extent of $0.1$
to $1''$ in typical high-redshift QSOs.
\end{abstract}


\section{INTRODUCTION}

  The origin of the ionizing background at high redshift has been a
long-standing question (Bechtold \etal 1987; Miralda-Escud\'e \&
Ostriker 1990; Songaila, Cowie, \& Lilly 1990; Madau 1991; Haardt \&
Madau 1996; Giroux \& Shapiro 1996). The possible sources include QSO's,
star-forming galaxies, and the cooling radiation from hot gas in halos.
One of the interests in measuring the intensity of the ionizing
background is that, with the present understanding of the \lya forest as
arising from the gravitational collapse of structure (see Rauch 1998 for
a review), the observed mean flux decrement in the \lya forest provides
a measurement of the parameter $\Omega_b^2/\Gamma$, where $\Omega_b$ is
the baryon density in units of the critical density and $\Gamma$ is the
photoionization rate due to the background (e.g., Rauch \etal 1997,
Weinberg \etal 1997, 1999; McDonald \etal 2000). Once $\Gamma$ is known
independently, the \lya forest provides a measurement of $\Omega_b$ at
low redshift which should agree with the values derived from Big Bang
nucleosynthesis (O'Meara \etal 2001, Pettini \& Bowen 2001) and from the
Cosmic Microwave Background spectrum of temperature fluctuations
(Netterfield \etal 2001, Pryke \etal 2001, Stompor \etal 2001), if our
cosmological ideas are correct.  

  The proximity effect (see Scott \etal 2000 and references therein),
consisting of the reduction of the number of \lya absorption lines in
the vicinity of a QSO due to its own ionizing radiation,
has been used to measure the intensity of the ionizing background, 
yielding a value $\Gamma\simeq 2\times 10^{-12}$. This
method is subject to several possible systematic errors, including
uncertainties in the QSO redshift, the effects of gravitational lensing
of the QSO (which makes it seem more luminous), QSO variability over
the photoionization timescale, or the clustering of gas around QSOs
which may partially balance the ionization effect.

  The simple counting of QSO's as a function of flux in the sky can give
us the emissivity of ionizing radiation, and by taking into account the
absorption by the intergalactic medium, which is also directly
determined in QSO spectra, we can calculate a lower limit to the
background intensity under the assumption that sources other than QSO's
are not important (assuming that the escape fraction of ionizing photons
from star-forming galaxies is negligible). This approach yields $\Gamma
\gtrsim 10^{-12}$, which implies a lower limit $\Omega_b h^2 > 0.02$
(Rauch \etal 1997; McDonald \etal 2000, 2001).

  In this paper, we examine if the ionizing radiation from QSOs could be
significantly absorbed by hydrogen in the halo where the QSO is located,
which normally produces the damped absorption systems, but is highly
photoionized in the presence of the QSO. Although absorption by Lyman
limit systems is taken into account when computing the intensity of the
ionizing background, (Haardt \& Madau 1996 and references therein), the
absorbers have always been assumed to be uncorrelated with the QSOs, so
any {\it self}-absorption arising in the halo of the QSO itself should
be in addition to the one computed from the general intergalactic gas.
Of course, any such absorption should be directly observable in the QSO
spectra; however, the self-absorption might be strong only in
low-luminosity QSOs, where the gas is less highly photoionized, and
surveys of Lyman limit systems have generally been done on the most
luminous QSOs.

  High redshift QSOs are also thought to be the sources that reionize
\heii in the intergalactic medium, which could have occurred as late as
$z=3$ (see Heap \etal 2000 and references therein). It is also of
interest to know if \heii ions in the QSO halos can produce significant
self-absorption of the \heii-ionizing radiation from QSOs to delay the
epoch at which the \heii reionization is completed.

  Another interesting consequence of the photoionization of gas in the
halo where the QSO resides is the extended narrow-line emission that
should be produced by the recombinations, as recently discussed by
Haiman \& Rees (2001). We will examine the predicted flux in \lya
emission.

\section{MODEL}

  Our model for the self-absorption in QSO's has two main parts. First,
each QSO of luminosity $L$ is assumed to be located in a Cold Dark
Matter (CDM) halo of mass $M$, with a unique relation $M(L)$, and a
probability $P_0$ that any halo will host a QSO. In other words, each
halo can be either in an active state, in which case the QSO luminosity
depends only on the halo mass, or in a quiescent state in which case
there is no QSO (in reality we expect a dispersion in this $M(L)$
relation; we will comment later on its effect on our results). Second, a
spherical model of the gas density as a function of radius is adopted
for each halo of mass $M$, which reproduces the observations of damped
\lya systems when the gas is all neutral. We then calculate the degree
of ionization of the gas in photoionization equilibrium in the presence
of the QSO flux, assuming that the density has not changed as a result
of the photoionization heating or hydrodynamic winds from the QSO.

  For the first part of the model, we can determine the relation $M(L)$
once we know the distributions of both $M$ and $L$. We adopt the
Press-Schecter formalism (Press \& Schecter 1974; Bond et al.\ 1991) for
the distribution of halo masses, with the CDM model with cosmological
constant with parameters $\Omega_{\Lambda}=0.7$, $\Omega_{m}=0.3$,
$\Omega_{b}=0.04$, and $h=0.7$. We use the parameter $\delta_{c}=1.69$
(with top-hat filter) for the threshold overdensity required to form a
halo, and the relations $M=(4\pi/3)\, 18\pi^{2}\rho_{crit} r_{vir}^{3}$,
and $V_{c}^{2}=GM/r_{vir}$ to relate halo mass, virial radius, and
circular velocity.

  For the QSO luminosity function (LF), we use the double power-law
model of Pei (1995):
\begin{equation}
\ \Phi(L_B; z) = {\Phi_{*}/L_{B*} \over
(L/L_{B*})^{\beta_{l}} + (L/L_{B*})^{\beta_{k}} } ~, 
\end{equation}
where $L_B$ is the B-band luminosity. All our results will be presented
at $z=3$, when $L_{B*}= 1.2 \times 10^{13}\, L_{B\odot}$, and
$\phi_{\star}= 619.25\, {\rm Gpc}^{-3}$ (from Table 1 in Pei 1995, after
correcting to our adopted cosmological model and Hubble constant). To
convert to the luminosity per unit frequency at the Lyman limit, we
convert to AB magnitude (see Oke 1974) using $M_B = 5.4 - 2.5
\log(L_B/L_{B\odot}) = M_{AB}(4400 {\rm \AA}) + 0.12$ (Schmidt,
Schneider, \& Gunn 1995), and we use a spectral index $\alpha= 0.5$ from
$\lambda=4400$ \AA\ to $\lambda= 1216$ \AA\ (where $L_{\nu} \propto
\nu^{-\alpha}$), and $\alpha= -1.77$ from $\lambda=1216$ \AA\ to
$\lambda = 912$ \AA\ (Zheng et al.\ 1997). This gives $L_{\nu *}=1.18
\times 10^{31}\ergs \, {\rm Hz}^{-1}$ at the Lyman limit.

  Figure 1 shows the relation obtained between the B-band QSO luminosity
and the halo mass, for four different values of $P_0$, by simply
requiring that the total number of halos with mass greater than $M$
times $P_0$ is equal to the total number of QSOs with luminosity greater
than $L$.

  For the density profile of the gas in each halo, we use the same
model as McDonald \& Miralda-Escud\'e (1999): a spherically symmetric
gas distribution with an exponential density profile,
\begin{equation}
\rho(r)=\rho_{0}\, \exp(-r/r_{g}) ~.
\label{densp}
\end{equation}
The two parameters of the mass distribution, the radius $r_g$ and the
density normalization $\rho_0$, need to be chosen to match the observed
column density distribution of the damped \lya systems, when the density
$\rho$ is assumed to be all neutral gas. We use the same parameters as
McDonald \& Miralda-Escud\'e (1999) at $z=3$ (see their Fig.3):
$c_{g}=r_{g}/r_{vir}=0.04$, and $\rho_0$ determined by a fraction of
the baryon mass in the halo in atomic gas form $f_{HI}=0.1$, both
assumed to be independent of halo mass. The resulting cumulative number
of absorbers above a column density $N_{HI}$ per unit redshift is shown
in Figure 2 as the thin line.
The points in this figure are the observations from Storrie-Lombardi
et al.\ (1996); to obtain these points, we divided the total number
of absorbers shown in their Figure 4 by their total redshift pathlength,
$\Delta z=74.7$ (which we obtain by adding the redshift pathlength for
their systems between $z=2$ and $z=3$, equal to $31.3$, and for their
systems at $z>3$, which is $43.5$, according to their Table 4).

  The calculation of the self-absorption column density is then done
simply by calculating the neutral fraction as a function of radius
for each halo mass $M$ and corresponding QSO luminosity $L$, with
the density profile in equation (\ref{densp}), computing the ionizing
flux from the QSO at each radius, and assuming photoionization
equilibrium with the recombination coefficient
$\alpha_A=4\times 10^{-13}\cm^{3}\, {\rm s} ^{-1}$ 
(at an assumed gas temperature $T=10^4\, {\rm K}$). 
        
\section{RESULTS}

  The predicted neutral hydrogen column density as a function of the
QSO luminosity is shown in Figure 3. The Lyman limit optical depth,
$\tau_{LL}$, reaches unity at a column density $N_{HI}=1.6\times 10^{17}
\cm^{-2}$, indicated by the thin solid line. In general, $\tau_{LL}$ is
predicted to be quite small. Even for $L=0.01 L_*$, $\tau_{LL}\simeq
0.1$ for $P_0=10^{-2}$. The predicted \heii column density is shown in
Figure 4. We have assumed a mean spectral index $L_{\nu}\propto
\nu^{-1.5}$ between the ionization edges of \hi and \heii, which implies
$N_{HeII}=13.4 N_{HI}$ (we do not include self-shielding effects).
Although the \heii Lyman limit opacity is higher than for \hi, it
still does not produce a very significant absorption.
Integrating the quantity $L\ e^{-\tau_{LL}}$ over the QSO luminosity
function, we find that the overall reduction in the emissivity is a
factor $(0.85, 0.94, 0.97, 0.99)$ for \hi, and $(0.71, 0.86, 0.93,
0.97)$ for \heii, for $P_0=10^{-1}, 10^{-2}, 10^{-3}, 10^{-4}$,
respectively. Notice that the reduction in the total ionization rate
from the ionizing background is smaller than these factors because
absorption is lower at frequencies above the Lyman limit.

  It is easy to see that in our model, the \hi column density is
proportional to the halo mass divided by the QSO luminosity. At fixed
$r/r_g$, the gas density is constant, and the flux is proportional to
$L/r_g^2$, so the neutral fraction goes as $r_g^2/L$, and the column
density goes as $r_g^3/L$. Since $r_g^3 \propto r_{vir}^3 \propto M$
(where $r_{vir}$ is the virial radius of the halo), we have
$N_{HI}\propto M/L$. If the halo properties were independent of QSO
luminosity, then of course $N_{HI}\propto L^{-1}$. Our curves show a
slower decrease of the column density with luminosity because of the
increasing halo mass. The column density also decreases with $P_0$
at fixed $L$ proportionally to the halo mass, which is shown in
Figure 1.
  
Lyman limit absorption at the QSO redshift which decreases with
luminosity. A possible difficulty is that, in photoionization
equilibrium, an equal number of recombinations and photoionizations
will take place, and about 38\% of the recombinations are direct to the
ground state and produce photons just above the Lyman limit
frequency, smoothing the discontinuity due to the absorption. As we
shall see below, the angular size of the damped systems from which
these recombination photons would come is probably in the range $0.1 -
1''$, and therefore difficult to resolve from the ground.

\section{DISCUSSION}

  We have computed both the neutral hydrogen and helium column densities
of self-absorption in QSOs in a model where the QSOs are located in the
same halos that produce the damped \lya absorption systems. The model
assumes that each QSO luminosity corresponds to a halo mass, and that
the gas in damped \lya systems that is ordinarily present in a halo in
the absence of a QSO is only photoionized, but not expelled when the QSO
is present.

  The parameter that we vary in our model is the probability $P_0$ that
a given halo has an active QSO. Large values of $P_0$ imply
long-lived QSOs. For a radiative efficiency $\epsilon$, the Salpeter
time for the growth of the black hole mass is $Mc^2/(\epsilon L_{Edd})
\gtrsim 4\times 10^8 \epsilon$ years. The typical time for the QSO
luminosity function to evolve is $\sim 10^9$ years; if we require that
the black hole mass function does not evolve on a shorter timescale, and
that $\epsilon < 0.1$, then $P_0 < 0.04$. From Figure 3, this limits the
self-absorption from the gas in damped \lya systems to $ \tau_{LL} <
0.02$ at $L_*$, and $\tau_{LL} < 0.2$ at $0.01 L_*$.

  An alternative to the assumption we made of a fixed relation between
QSO luminosity and halo mass is that, even if QSOs are located in halos,
there is little correlation between luminosity and halo mass. In this
case, we can define a minimum halo mass $M_{min}$ that hosts QSOs above
a certain luminosity $L_{min}$, with probability $P_0$; then, Figures 3
and 4 still give the HI  and HeII column densities expected at
$L_{min}$, but at higher luminosities the average column density would
decrease as $N_{HI} \propto L^{-1}$, more rapidly than in
Figures 3 and 4.

  We note here a possible caveat of our model when $P_0 \gtrsim
10^{-2}$: the halos that account for most of the observed damped \lya
systems have velocity dispersions
in the range $40$ to $150 \kms$, or, at $z=3$, total masses of $10^{10}$
to $10^{12} \msun$ (e.g., Gardner \etal 1997). From Figure 1, this mass
range corresponds to rather low QSO luminosities if $P_0\sim 0.01$. This
means that the absorption in QSOs of luminosities $\gtrsim 0.1 L_*$ is
in this case due to halos with masses as large as $10^{13} \msun$, which
could have different physical conditions than the more numerous
lower-mass halos that are mostly responsible for damped \lya absorbers.

  The Lyman limit self-absorption shown in Figures 3 and 4 could be
substantially increased if the absorbing gas is clumpy. We have assumed
that the gas density in the halo is smoothly distributed, following the
density profile in equation (\ref{densp}). However, the observations of
multiple metal absorption lines associated with damped \lya systems
(e.g., Prochaska \& Wolfe 1997) show that the gas is actually clumpy.
The typical number of multiple absorbers that are observed tells us that
the covering factor of these clumps is of order unity, but the clumping
factor depends on the size of the clumps. The clumping factor may be not
much larger than unity in a scenario like that proposed by Haehnelt,
Steinmetz, \& Rauch (1998), where the clumps are due to halo mergers and
the complicated line structure arise in a continuous medium of halo gas
from velocity caustics and moderate density variations. But the clumping
factor could be high if the clumps were much smaller than the overall
size of the damped \lya systems. The absorbing column density increases
proportionally to the clumping factor.

  Another possible modification of the model we have used here is that
the gas observed in damped \lya systems does not exist in every halo,
but only in a certain fraction $f$ of halos. In order to preserve the
rate of incidence of the observed absorption systems, we then need to
increase the radius $r_g$ in equation (\ref{densp}) by a factor
$f^{-1/2}$, and decrease the central density by $f^{1/2}$ to have the
same column density distribution in the absence of ionization. With the
same QSO luminosity, the HI column densities in the presence of the QSO
are increased by a factor $f^{-1/2}$. Thus, making the damped \lya
absorbers bigger and less abundant would increase the amount of
self-absorption, if QSOs are located in the halos that contain the
damped \lya systems. This could be the case, for example, if QSO activity
usually takes place in recently merged halos with a lot of fresh gas.

\subsection{Narrow-line emission}

  The absorption of the continuum ionizing radiation of a faint QSO by
gas with velocity dispersions similar to those in damped \lya systems
implies that narrow \lya emission lines should be observed, containing
about a third of the energy that is absorbed in Lyman limit photons.
This emission line could remain unresolved, since damped \lya systems
are small. For example, in a halo of mass $10^{12}$ $\msun$,
$r_g= 3.15 \kpc$ (see eq.\ [\ref{densp}]) at $z=3$, corresponding to an
angular size of $0.41''$. This narrow emission line would be superposed
with the common broad absorption lines in luminous QSOs.

  The \lya emission from halo gas around QSOs was recently considered
by Haiman \& Rees (2001). Their model assumes that all the baryons in a
CDM halo are in an extended gas halo containing a hot and cold
phase, with a large fraction of the baryons in the cold phase in most
halos of interest (see their Fig. 1). In our model, with our choice
of parameters $f_{HI}=0.1$ and $r_g/r_{vir}=0.04$, only 10\% of the
baryons are in the gaseous halo, but these are mostly concentrated
within $\sim 10\%$ of $r_{vir}$. Since the gas density profile in the
model of Haiman \& Rees is roughly isothermal, their prediction for the
abundance of damped \lya systems in the absence of a central source
should not be very different from that of our model, which we have
shown agrees with the observations.


  Generally, narrow \lya emission may arise from gas over a wide range
of radius and densities, not only associated with the damped absorption
systems but with lower column density absorbers as well.
The exponential gas density profile in equation (\ref{densp}) for our
model is intended to approximate the effects of self-shielding when
computing the number of damped systems in the absence of a central
photoionizing source (when the photoionization is due to the external
background), which should cause the neutral density to drop sharply
outside the self-shielded region. But the total gas density is likely
to drop more slowly. As an example, if $\rho_g\propto r^{-2}$, and the
gas is mostly ionized, then the recombination rate per unit volume is
proportional to $r^{-4}$, and the \lya surface brightness drops as
$r^{-3}$. This `` \lya fuzz'' should always be present in all halos as
long as the gas is ionized, whether or not a central QSO is present.
In the absence of a central QSO, the \lya emission should have a core
at the radius where the gas becomes self-shielded against the external
background, or in other words, the radius at which Lyman limit
absorption would be seen against a background source. When the QSO
turns on, the gas within this core becomes ionized, and the steep
power-law surface brightness profile can be extended inwards, making
it much brighter. However, since Lyman limit systems are only $\sim$
10 times more abundant than damped \lya systems, their typical extent
in halos should be only $\sim 3$ times larger in radius than the damped
systems, which as mentioned earlier would be hardly resolved from the
ground. Therefore, the presence of a QSO can only increase the emitted
\lya in a central region of size $\sim 1''$, except in very massive and
gas rich halos where the self-shielded region might be larger.

  The \lya surface brightness predicted by our model can be estimated
from the \hi absorption column densities shown in Figure 3. Let us take,
for example, the curves for $P_0=10^{-2}$ in Figures 1 and 3, and
consider the case of a halo with $\sigma=150 \kms$, which implies a
mass $10^{12} \msun$ at $z=3$. Our model associates this halo with a QSO
of $L_B\simeq 0.03 L_*\simeq 10^{29.5} \ergs\, {\rm Hz}^{-1}$, which has
a Lyman limit absorption $\tau_{LL}\simeq 0.04$ from Figure 3. The
frequency over which $\tau_{LL}$ decreases is $\Delta \nu\sim 10^{15}$
Hz, so the power absorbed in hydrogen photoionizations is about
$10^{43} \ergs$, and the power radiated in \lya photons is
$10^{42.5} \ergs$, which at $z=3$ produces a \lya flux $\sim 10^{-16}
\ergs\cm^{-2}$. If this flux is extended with a profile falling as
$r^{-3}$, with a core of $\sim 0.5$ arc seconds (corresponding to the
radius $r_g$ of the damped absorption system), the surface brightness
would be $(10^{-17.5}, 10^{-19.5} \ergs\cm^{-2}\, {\rm arcsec}^{-2}$
at angular separations of $(0.5, 3)''$, respectively.
Just like the \hi absorption column density, this \lya surface
brightness can be increased by the clumping factor of the gas.
In comparison, from Figure 2 of Haiman \& Rees, they conclude that a
similar halo (with $T_{vir}=2\times 10^6$ K) would produce a \lya
surface brightness of $2\times 10^{-16}$ Hz at 3''. The large difference
with our prediction of the \lya surface brightness can be traced to the
high clumping factor of the cold gas in the model of Haiman \&
Rees (see their eq. 1), and to their assumption that the \lya flux
comes mostly from a large region, close to the virial radius of the
halo, whereas we assume it is much more concentrated to the center.

  Recently, Steidel \etal (2000) have reported the discovery of two
``\lya blobs'' of emission, with surface brightness
$\sim 10^{-17} \ergs\cm^{-2}$, and angular scale of $\sim 10''$. While
the surface brightness is similar to the expected value quoted above
from a typical damped absorber once the gas is ionized, the angular
scale is very large. These systems are therefore likely produced in
exceptionally massive and gas-rich halos.

  To summarize, if extended gas is present in the halos where QSOs
are located, with a similar distribution as the average profiles
derived from the observed damped \lya systems, then we should observe
narrow-line \lya emission, extended over a region comparable to the
size of damped \lya systems, which is typically less than $1''$. As
mentioned at the end of \S 3, there should also be extended emission
of the Lyman continuum photons from direct recombinations to the
ground state. We showed in Figures 3 and 4 that the absorption column
densities expected are optically thin even for very low QSO
luminosities; therefore, for fixed halo properties, the \lya
fuzz should be easier to observe around fainter QSOs,
which already produce sufficient flux to ionize all the halo gas,
implying a \lya brightness independent of QSO luminosity
(our Fig. 3 shows that $N_{HI}$ decreases more slowly than
$L^{-1}$ with luminosity, which implies an increasing \lya surface
brightness with luminosity, only because of the assumption in our
model that brighter QSOs are located in more massive halos which
contain more gas). 
If the absorption or the associated \lya emission are not observed at
the predicted level, the conclusion should be that either QSOs are
in halos that do not contain the average amount of gas that is inferred
from the abundance of absorption
systems, or the QSOs themselves have expelled this gas in winds.

 We would like to acknowledge Patrick McDonald for many stimulating discussions.

{}

\clearpage

{ \begin{figure}[t]
\vspace{2.5in}
\label{fig1}
\includegraphics{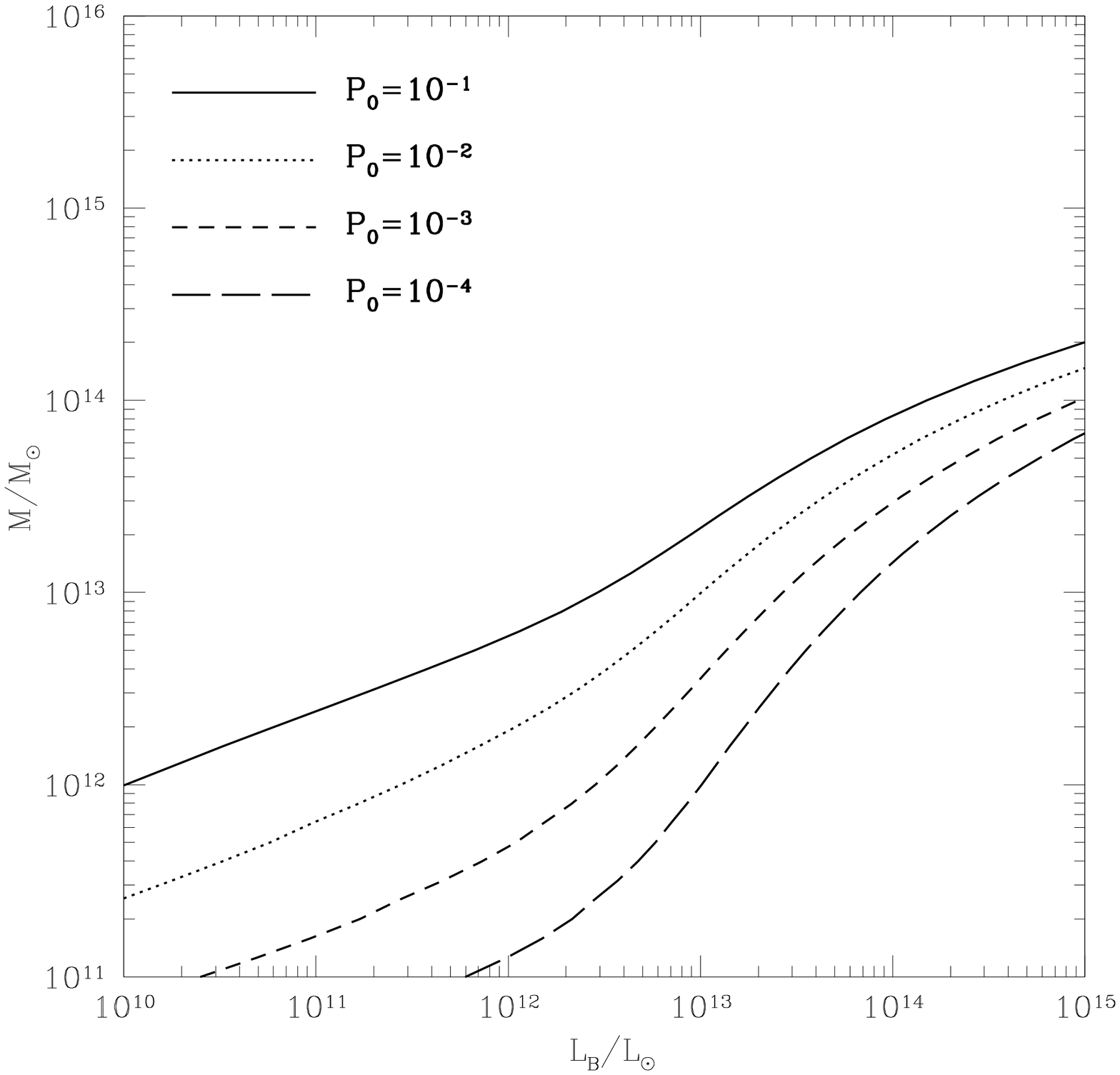}
\figcaption{Relation between halo mass and QSO luminosity for different
probabilities $P_0$.
\label{fig1}}
\end{figure}}

\clearpage

{\begin{figure}[t]
\vspace{2.5in}
\label{fig2}
\includegraphics{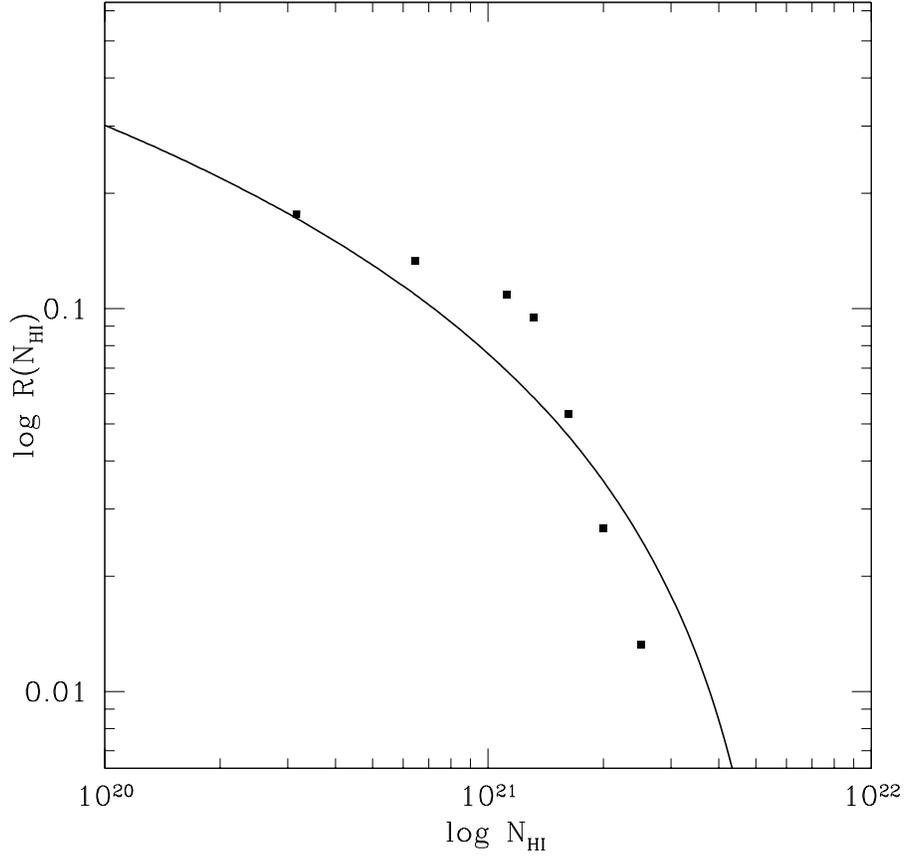}
\figcaption{Cumulative number of absorbers per unit redshift as a
function of \hi column density. Squares are the observed values from
Storrie-Lombardi et al. (1996); the thin solid line is the prediction
of our model.
\label{Fig2}}
\end{figure}}

\clearpage

{\begin{figure}[t]
\vspace{2.5in}
\label{fig3}
\includegraphics{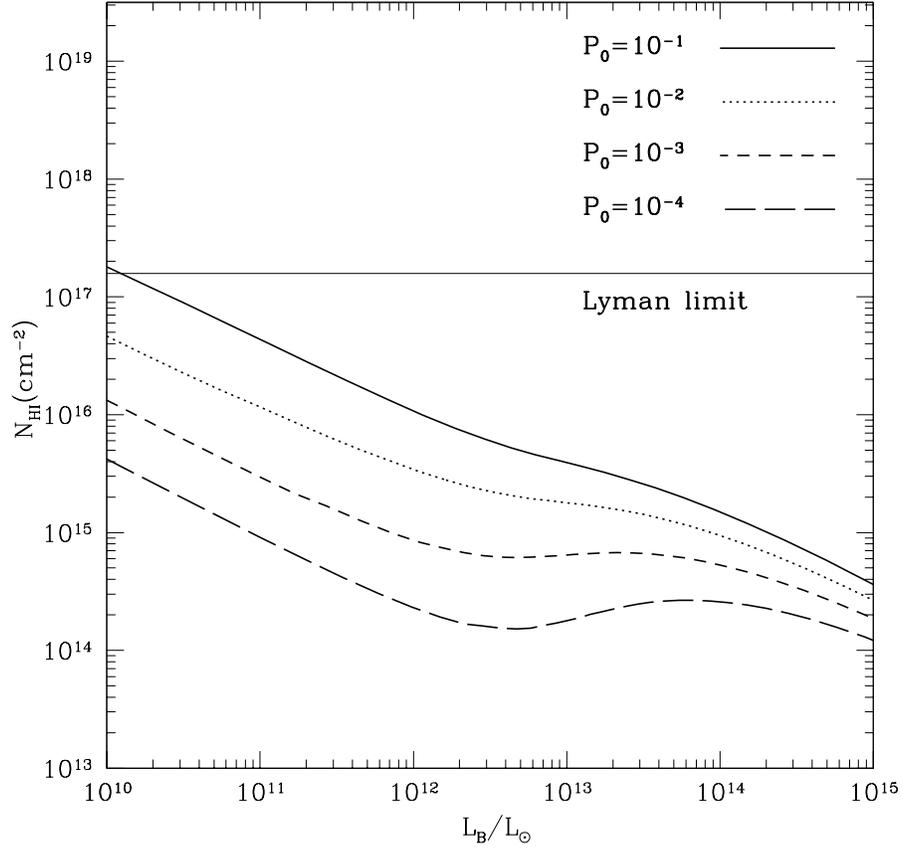}
\figcaption{Neutral hydrogen column density as a function of QSO B-band
luminosity for different probabilities $P_0$. The thin horizontal line
indicates a Lyman limit optical depth of unity.
\label{fig3}}
\end{figure}}

\clearpage

{\begin{figure}[t]
\vspace{2.5in}
\label{fig4}
\includegraphics{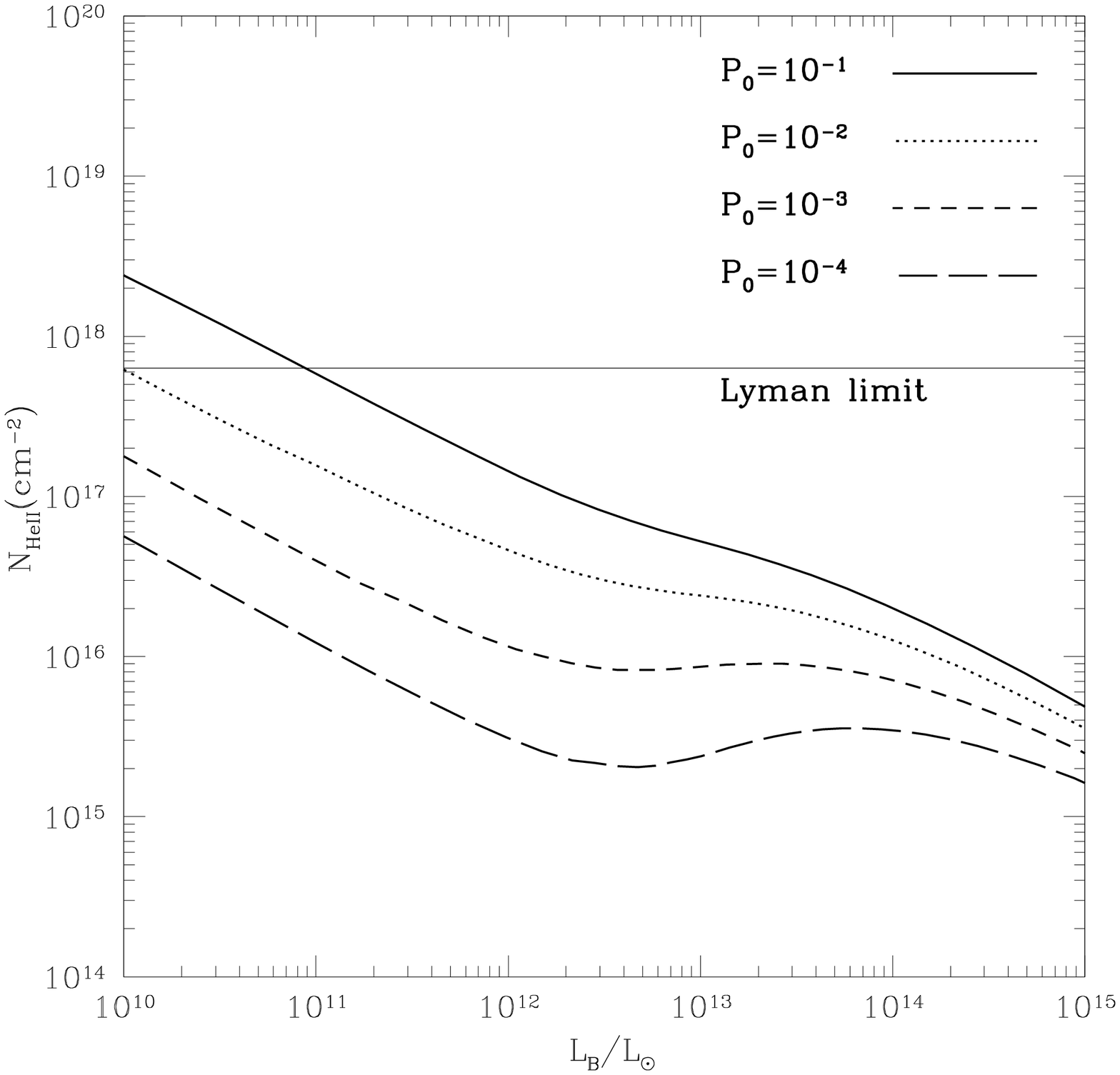}
\figcaption{\heii column density as a function of QSO B-band luminosity
for different probabilities $P_0$. The thin horizontal line indicates a
Lyman limit optical depth of unity.
\label{fig4}}
\end{figure}}

\end{document}